\documentclass[prd,letterpaper,onecolumn,superscriptaddress,nofootinbib]{revtex4}
\usepackage{amsmath,amssymb,epsfig,natbib,bm,psfrag}

\begin{document}

\newcommand{\ud}{{\text{d}}}
\newcommand{\vnhat}{\hat{\mathbf{n}}}
\newcommand{\valpha}{\boldsymbol{\alpha}}
\newcommand{\vgrad}{\boldsymbol{\nabla}}
\newcommand{\vk}{{\mathbf{k}}}
\newcommand{\vl}{{\mathbf{l}}}
\renewcommand{\vr}{{\mathbf{r}}}
\newcommand{\vL}{{\mathbf{L}}}
\newcommand{\vq}{{\mathbf{q}}}
\newcommand{\vqp}{{{\mathbf{q}}_+}}
\newcommand{\vqm}{{{\mathbf{q}}_-}}
\newcommand{\half}{{\textstyle{\frac{1}{2}}}}
\newcommand{\third}{{\textstyle{\frac{1}{3}}}}
\newcommand{\twothirds}{{\textstyle{\frac{2}{3}}}}
\newcommand{\Cgl}{C_{\text{gl}}}
\newcommand{\Cgltwo}{C_{\text{gl},2}}

\newcommand{\primvar}{\mathcal{R}}

\newcommand{\Mpc}{\text{Mpc}}
\newcommand{\numfrac}[2]{{\textstyle \frac{#1}{#2}}}
\newcommand{\ra}{\rangle}
\newcommand{\la}{\langle}
\renewcommand{\d}{\text{d}}
\newcommand{\grad}{\nabla}

\newcommand{\begm}{\begin{pmatrix}}
\newcommand{\enm}{\end{pmatrix}}

\newcommand{\threej}[6]{{\begm #1 & #2 & #3 \\ #4 & #5 & #6 \enm}}
\newcommand{\fsky}{f_{\text{sky}}}

\newcommand{\cla}{\mathcal{A}}
\newcommand{\clb}{\mathcal{B}}
\newcommand{\clc}{\mathcal{C}}
\newcommand{\cle}{\mathcal{E}}
\newcommand{\clf}{\mathcal{F}}
\newcommand{\clg}{\mathcal{G}}
\newcommand{\clh}{\mathcal{H}}
\newcommand{\cli}{\mathcal{I}}
\newcommand{\clj}{\mathcal{J}}
\newcommand{\clk}{\mathcal{K}}
\newcommand{\cll}{\mathcal{L}}
\newcommand{\clm}{\mathcal{M}}
\newcommand{\cln}{\mathcal{N}}
\newcommand{\clo
}{\mathcal{O}}
\newcommand{\clp}{\mathcal{P}}
\newcommand{\clq}{\mathcal{Q}}
\newcommand{\clr}{\mathcal{R}}
\newcommand{\cls}{\mathcal{S}}
\newcommand{\clt}{\mathcal{T}}
\newcommand{\clu}{\mathcal{U}}
\newcommand{\clv}{\mathcal{V}}
\newcommand{\clw}{\mathcal{W}}
\newcommand{\clx}{\mathcal{X}}
\newcommand{\cly}{\mathcal{Y}}
\newcommand{\clz}{\mathcal{Z}}
\newcommand{\CMBFAST}{\textsc{cmbfast}}
\newcommand{\CAMB}{\textsc{camb}}
\newcommand{\RECFAST}{\textsc{recfast}}
\newcommand{\COSMOMC}{\textsc{CosmoMC}}
\newcommand{\Healpix}{\textsc{healpix}}
\newcommand{\HALOFIT}{\textsc{halofit}}
\newcommand{\Omtot}{\Omega_{\mathrm{tot}}}
\newcommand{\Omb}{\Omega_{\mathrm{b}}}
\newcommand{\Omc}{\Omega_{\mathrm{c}}}
\newcommand{\Omm}{\Omega_{\mathrm{m}}}
\newcommand{\omb}{\omega_{\mathrm{b}}}
\newcommand{\omc}{\omega_{\mathrm{c}}}
\newcommand{\omm}{\omega_{\mathrm{m}}}
\newcommand{\Omdm}{\Omega_{\mathrm{DM}}}
\newcommand{\Omnu}{\Omega_{\nu}}

\newcommand{\Oml}{\Omega_\Lambda}
\newcommand{\OmK}{\Omega_K}

\newcommand{\Hunit}{~\text{km}~\text{s}^{-1} \Mpc^{-1}}
\newcommand{\Gyr}{{\rm Gyr}}

\newcommand{\nrun}{n_{\text{run}}}

\newcommand{\lmax}{l_{\text{max}}}

\newcommand{\zre}{z_{\text{re}}}
\newcommand{\mpl}{m_{\text{Pl}}}

\newcommand{\vphi}{\mathbf{\phi}}
\newcommand{\vv}{\mathbf{v}}
\newcommand{\vd}{\mathbf{d}}
\newcommand{\vC}{\mathbf{C}}
\newcommand{\vT}{\mathbf{T}}
\newcommand{\vX}{\mathbf{X}}
\newcommand{\vn}{\mathbf{n}}
\newcommand{\vy}{\mathbf{y}}
\newcommand{\mN}{\bm{N}}
\newcommand{\eV}{\,\text{eV}}
\newcommand{\vtheta}{\bm{\theta}}
\newcommand{\tT}{\tilde{T}}
\newcommand{\tE}{\tilde{E}}
\newcommand{\tB}{\tilde{B}}

\newcommand{\mCh}{\hat{\bm{C}}}
\newcommand{\Ch}{\hat{C}}

\newcommand{\Bt}{\tilde{B}}
\newcommand{\Et}{\tilde{E}}
\newcommand{\bld}[1]{\mathrm{#1}}
\newcommand{\mLambda}{\bm{\Lambda}}
\newcommand{\mA}{\bm{A}}
\newcommand{\mC}{\bm{C}}
\newcommand{\mQ}{\bm{Q}}
\newcommand{\mU}{\bm{U}}
\newcommand{\mX}{\bm{X}}
\newcommand{\mV}{\bm{V}}
\newcommand{\mP}{\bm{P}}
\newcommand{\mR}{\bm{R}}
\newcommand{\mW}{\bm{W}}
\newcommand{\mD}{\bm{D}}
\newcommand{\mI}{\bm{I}}
\newcommand{\mH}{\bm{H}}
\newcommand{\mM}{\bm{M}}
\newcommand{\mS}{\bm{S}}
\newcommand{\mzero}{\bm{0}}
\newcommand{\mL}{\bm{L}}

\newcommand{\btheta}{\bm{\theta}}
\newcommand{\bphi}{\bm{\psi}}

\newcommand{\vb}{\mathbf{b}}
\newcommand{\vA}{\mathbf{A}}
\newcommand{\vAt}{\tilde{\mathbf{A}}}
\newcommand{\ve}{\mathbf{e}}
\newcommand{\vE}{\mathbf{E}}
\newcommand{\vB}{\mathbf{B}}
\newcommand{\vEt}{\tilde{\mathbf{E}}}
\newcommand{\vBt}{\tilde{\mathbf{B}}}
\newcommand{\vEw}{\mathbf{E}_W}
\newcommand{\vBw}{\mathbf{B}_W}
\newcommand{\vx}{\mathbf{x}}
\newcommand{\vXt}{\tilde{\vX}}
\newcommand{\vXb}{\bar{\vX}}
\newcommand{\vTb}{\bar{\vT}}
\newcommand{\vTt}{\tilde{\vT}}
\newcommand{\vY}{\mathbf{Y}}
\newcommand{\vBwr}{{\vBw^{(R)}}}
\newcommand{\RW}{{W^{(R)}}}

\newcommand{\mUt}{\tilde{\mU}}
\newcommand{\mVt}{\tilde{\mV}}
\newcommand{\mDt}{\tilde{\mD}}

\newcommand{\Rot}{\begm \mzero &\mI \\ -\mI & \mzero \enm}
\newcommand{\Pt}{\begm \vEt \\ \vBt \enm}

\newcommand{\edth}{\,\eth\,}
\renewcommand{\beth}{\,\overline{\eth}\,}

\newcommand{\xil}{\tilde{\xi}}

\newcommand{\sE}{{}_{|s|}E}
\newcommand{\sB}{{}_{|s|}B}
\newcommand{\sElm}{\sE_{lm}}
\newcommand{\sBlm}{\sB_{lm}}
\newcommand{\angarg}{}


\title{Lensed CMB power spectra from all-sky correlation functions}

\author{Anthony Challinor}
 \email{a.d.challinor@mrao.cam.ac.uk}
 \affiliation{Astrophysics Group, Cavendish Laboratory, Madingley Road,
Cambridge CB3 0HE, U.K.}

\author{Antony Lewis}
 \homepage{http://cosmologist.info}
 \affiliation{CITA, 60 St. George St, Toronto M5S 3H8, ON, Canada.}

\begin{abstract}
Weak lensing of the CMB changes the unlensed temperature anisotropy and
polarization power spectra. Accounting for the lensing effect will be
crucial to obtain accurate parameter constraints from sensitive CMB
observations.
Methods for computing the lensed power
spectra using a low-order perturbative expansion are not good enough
for percent-level accuracy. Non-perturbative flat-sky methods are more
accurate, but curvature effects change the spectra at the $0.3$--$1\%$ level.
We describe a new, accurate and fast, full-sky
correlation-function method for computing the lensing effect on CMB power
spectra to better
than $0.1\%$ at $l\alt 2500$ (within the approximation that the lensing
potential is linear and Gaussian). We also discuss the effect of non-linear evolution of the gravitational potential on the lensed power spectra.
Our fast numerical code is publicly
available.
\end{abstract}


\maketitle

\section{Introduction}

The CMB temperature and polarization anisotropies are being measured
with ever increasing precision. The statistics of the anisotropies
already provide valuable limits on cosmological parameters, as well as
constraints on early-universe physics. As we enter the era of precision
measurement, with signal-dominated observations out to small angular scales,
non-linear effects will become increasingly important.
One of the most significant of these over scales of most interest
for parameter estimation is weak gravitational lensing by large scale
structure. Fortunately it can be
modelled accurately as a second-order effect: the linear gravitational
potential along the line of sight lenses the linear perturbations at the last
scattering surface (see e.g.\ Refs.~\cite{Seljak:1996ve,
Zaldarriaga:1998ar,Hu:2000ee} and references therein).
Modelling of fully non-linear evolution is not required for the near
future on scales of several arcminutes (corresponding to multipoles
$l \alt 2000$) for the temperature and electric polarization power spectra.
Non-linear corrections can easily be applied to the lensing potential
if and when required, provided that its non-Gaussianity can be
ignored~\cite{Seljak:1996ve}.

In principle, the weak-lensing contribution to the observed sky can
probably be subtracted given sufficiently accurate and clean high-resolution
observations. Early work in this
area~\cite{Seljak:1998aq,Guzik:2000ju,Hu:2001tn,Hu:2001kj}
suggested a limit on the
accuracy of this reconstruction due to the statistical nature of the
(unknown) unlensed CMB fields. More recently, it has been argued that
polarization removes this limit in models where lensing
is the only source of $B$-mode polarization on small
scales~\cite{Hirata:2003ka}. If subtraction could be done exactly
we could recover the unlensed Gaussian sky, and use this for all
further analysis. However current methods for subtracting the lensing
contribution are approximate, and not easy to apply to realistic
survey geometries. The result of imperfect lensing subtraction is
a sky with complicated, non-Gaussian statistics of the signal, and
significantly more complicated noise properties than the original (lensed)
observations. For observations in the near future, a
much simpler method to account for the lensing effect is to work with
the lensed sky itself, modelling the lensing effect by the expected change
in the power spectra and their covariances. The effects of lensing
non-Gaussianities on the covariance of the temperature and $E$-mode
polarization power spectra are likely to be small, but this will not be the
case for the $B$-mode spectrum once thermal-noise levels permit imaging
of the lens-induced $B$ modes~\cite{Smith:2004up}.
In this paper we discuss how to compute the lensed power spectra accurately.
The simulation of lensed skies
and the effect on parameter estimation is discussed in Ref.~\cite{Lewis:2005tp}.

On scales where the non-Gaussianity of the lensing potential can be ignored,
the calculation of the lensed power spectra is straightforward in principle.
However, achieving good accuracy on both large and small scales for all the CMB
observables is surprisingly difficult. The lensing action on the
CMB fields at scales approaching the r.m.s.\ of the lensing deflection angle
($\sim 3\, \mathrm{arcmin}$) cannot be accurately described
with a first-order Taylor expansion, as in the full-sky harmonic method of
Ref.~\cite{Hu:2000ee}.
There is not much power in the unlensed CMB on such scales, but a first-order Taylor expansion still gives lensed power spectra that
are inaccurate at the percent level for $l \agt 1000$.
The lensed CMB on scales well below the diffusion scale
is generated by the action of small-scale weak lenses
on the (relatively) large-scale unlensed CMB,
and a Taylor expansion should become more accurate
again~\cite{Seljak00lensrecon}. (However, non-linear effects
are also important on such scales.)
The breakdown of the Taylor expansion
can be easily fixed by using the flat-sky
correlation-function methods of Refs.~\cite{Seljak:1996ve,Zaldarriaga:1998ar},
which can handle the dominant effect of the lensing displacement in
a non-perturbative manner. However, a new problem then arises on scales
where the flat-sky approximation is not valid. As noted in
Ref.~\cite{Hu:2000ee}, this is not confined to large scales due to the
mode-coupling nature of lensing: degree-scale
lenses contribute significantly to the lensed power over a wide range of
observed scales. In this paper we develop a new method for computing the
lensed power spectra that is accurate on all scales where non-Gaussianity due to non-linear
effects is not important. We do this by calculating the lensed correlation
functions on the spherical sky. This allows us to include both
the non-perturbative effects of displacing small-scale CMB fluctuations,
and the effects of sky curvature.

This paper is arranged as follows. We start in
Section~\ref{sec:lensing} with a brief introduction to CMB
lensing, then in Section~\ref{sec:correlation} we review previous work on
flat-sky correlation-function methods and present our new
full-sky method and results. In Section~\ref{sec:comparison} we
compare our new results with the flat-sky correlation-function results of
Refs.~\cite{Seljak:1996ve,Zaldarriaga:1998ar} and the perturbative
harmonic result of Ref.~\cite{Hu:2000ee}, and explain why the latter
is not accurate enough for precision cosmology. The effect of non-linear
evolution of the density field on the lensed power spectra is considered in
Section~\ref{sec:nonlin}. We end with
conclusions, and include some technical results in the appendices.

\section{CMB lensing}
\label{sec:lensing}

Gradients in the gravitational potential transverse to the line of sight to
the last scattering surface cause deviations in the photon
propagation, so that points in a direction $\vnhat$ actually come
from points on the last scattering surface in a displaced direction
$\vnhat'$. Denote the lensed CMB temperature by $\tilde{\Theta}(\vnhat)$ and the unlensed
temperature by $\Theta(\vnhat)$, so the lensed field is given by
$\tilde{\Theta}(\vnhat) = \Theta(\vnhat')$.
The change in direction on the sky can be described by a
displacement vector field $\valpha(\vnhat) \equiv \vgrad \psi$, so
that (symbolically)
$\vnhat' = \vnhat +  \vgrad \psi$. Here $\psi$ is the lensing
potential which
encapsulates the deviations caused by potentials along the line of
sight. More rigorously, on a unit sphere the point $\vnhat'$ is obtained from
$\vnhat$ by moving a distance $|\vgrad\psi|$ along a geodesic in
the direction of $\vgrad\psi(\vnhat)$, where $\vgrad$ is the covariant
derivative on the sphere~\cite{Challinor02}.
We assume that the lensing is weak, so that the potentials may be
evaluated along the unperturbed path (i.e. we use the Born
approximation).  Lensing deflections are a few arcminutes, but are coherent over
degree scales, so this is a good approximation.

In terms of the zero-shear acceleration
potential $\Psi$, the lensing potential in a flat universe with
recombination at conformal distance $\chi_*$ is given by
the line-of-sight integral
\begin{equation}
\psi(\vnhat) = -2 \int_0^{\chi_*} \ud \chi\, \Psi(\chi \vnhat; \eta_0 -\chi)
\frac{\chi_* - \chi}{\chi \chi_*}.
\label{eq:1}
\end{equation}
Here we neglect the very small effect of late-time sources, including
reionization, and approximate recombination as instantaneous so that
the CMB is described by a single source plane at $\chi = \chi_*$. The
quantity $\eta_0 -\chi$ is the conformal time at which the photon was
at position $\chi \vnhat$.
With the Fourier convention
\begin{equation}
\Psi(\vx;\eta) = \int \frac{\ud^3 \vk}{(2\pi)^{3/2}}\,
\Psi(\vk;\eta) e^{i \vk \cdot \vx},
\label{eq:2}
\end{equation}
and power spectrum
\begin{equation}
\langle \Psi(\vk;\eta) \Psi^*(\vk';\eta') \rangle =
\frac{2\pi^2}{k^3} \clp_\Psi(k;\eta,\eta') \delta(\vk-\vk'),
\label{eq:3}
\end{equation}
the angular power spectrum of the lensing potential $\psi$ evaluates to
\begin{equation}
C_l^\psi = 16\pi \int \frac{\ud k}{k}\, \int_0^{\chi_*} \ud \chi\,
\int_0^{\chi_*} \ud \chi'\, \clp_\Psi(k;\eta_0-\chi,\eta_0-\chi')
j_l(k\chi) j_l(k\chi') \left(\frac{\chi_*-\chi}{\chi_*\chi}\right)
\left(\frac{\chi_*-\chi'}{\chi_*\chi'}\right).
\label{cpsi}
\end{equation}

\begin{figure}
\begin{center}
\psfig{figure=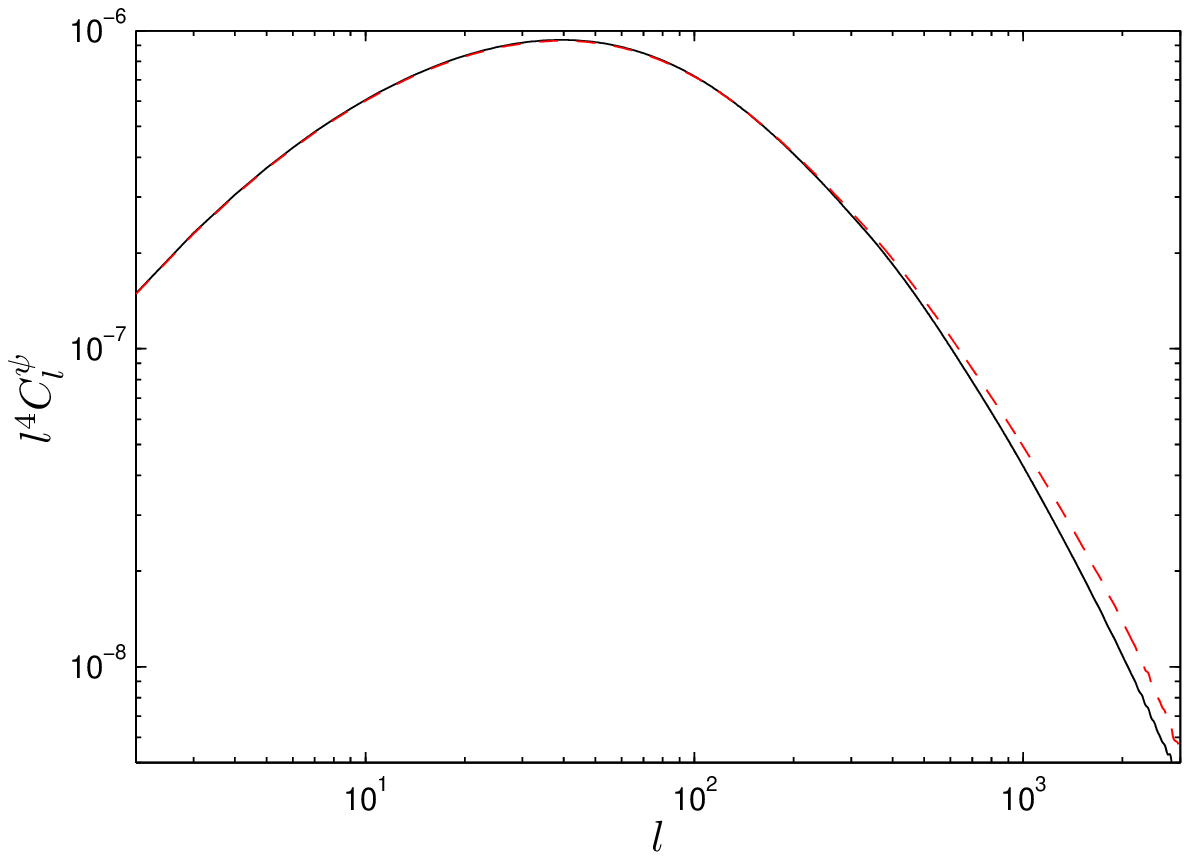,width=10cm}
\caption{The power spectrum of the lensing potential for a concordance $\Lambda$CDM model.
The linear theory spectrum (solid) is compared with the same model including non-linear corrections (dashed) from \HALOFIT~\cite{Smith:2002dz} using Eq.~\eqref{Tnonlin}.
\label{CPhi}}
\end{center}
\end{figure}

In linear theory we can define a transfer function $T_\Psi(k,\eta)$ so that $\Psi(\vk;\eta) = T_\Psi(k,\eta) \primvar(\vk)$ where $\primvar(\vk)$ is the
primordial comoving curvature perturbation (or other variable for isocurvature modes). We then have
\begin{equation}
C_l^\psi = 16\pi \int \frac{\ud k}{k}\, \clp_{\primvar}(k) \left[\int_0^{\chi_*} \ud \chi\,
T_\Psi(k;\eta_0-\chi) j_l(k\chi) \left(\frac{\chi_*-\chi}{\chi_*\chi}\right)\right]^2
\label{cpsi_transfer}
\end{equation}
where the primordial power spectrum is $\clp_{\primvar}(k)$. This can be computed easily numerically using \CAMB\footnote{\url{http://camb.info}}~\cite{Lewis:1999bs}, and a typical spectrum is shown in Fig.~\ref{CPhi}.

\section{Lensed correlation function}
\label{sec:correlation}

\subsection{Flat-sky limit}
\label{subsec:flat}

\begin{figure}
\begin{center}
\psfig{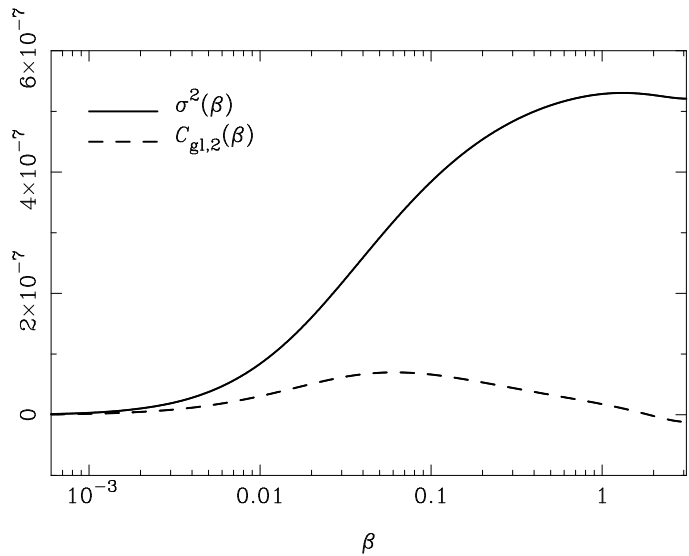}
\caption{
The functions $\sigma^2(\beta)\equiv \Cgl(0)-\Cgl(\beta)$ [solid] and $\Cgltwo(\beta)$ [dashed]
as a function of angular separation $\beta$ (in radians) for a typical
concordance model. The results are
calculated using the full-sky definitions of Eqs.~\eqref{cgl_def},
and use the linear power spectrum for $C_l^\psi$.
\label{sigmaplot}}
\end{center}
\end{figure}

We start by calculating the lensed correlation function in the flat-sky
limit, broadly following the method of Ref.~\cite{Seljak:1996ve}.
We use a 2D Fourier transform of the temperature field
\begin{equation}
\Theta(\vx) = \int \frac{\ud^2 \vl}{2\pi}\, \Theta(\vl) e^{i\vl\cdot \vx},
\label{eq:6}
\end{equation}
and the power spectrum for a statistically isotropic field is then
\begin{equation}
\langle \Theta(\vl) \Theta^*(\vl') \rangle = C_l^\Theta \delta(\vl-\vl').
\label{eq:7}
\end{equation}
Lensing re-maps the temperature according to
\begin{equation}
\tilde{\Theta}(\vx) = \Theta(\vx + \valpha),
\label{eq:8}
\end{equation}
where in linear theory the displacement vector $\valpha$ is a Gaussian field. We shall require
its correlation tensor $\langle \alpha_i(\vx) \alpha_j (\vx') \rangle$
to compute the lensed CMB power spectrum.
Introducing the Fourier transform of the
lensing potential, $\psi(\vl)$, we have
\begin{equation}
\valpha(\vx) = i \int \frac{\ud^2 \vl}{2\pi}\, \vl \psi(\vl) e^{i\vl\cdot \vx},
\label{eq:9}
\end{equation}
so that
\begin{equation}
\langle \alpha_i(\vx) \alpha_j(\vx') \rangle =
\int \frac{\ud^2 \vl}{(2\pi)^2} l_i l_j C_l^\psi e^{i\vl\cdot
(\vx-\vx')}.
\label{eq:10}
\end{equation}
By symmetry, the correlator can only depend on $\delta_{ij}$ and the trace-free
tensor $r_{\langle i} r_{j \rangle}$, where $\vr \equiv \vx - \vx'$.
Evaluating the coefficients of these two terms by taking the trace of
the correlator, and its contraction with $r^i r^j$, we find
\begin{equation}
\langle \alpha_i(\vx) \alpha_j(\vx') \rangle = \frac{1}{4\pi}
\int \ud l\, l^3 C_l^\psi J_0(lr) \delta_{ij} - \frac{1}{2\pi}
\int \ud l\, l^3 C_l^\psi J_2(lr) \hat{r}_{\langle i} \hat{r}_{j \rangle} ,
\label{eq:11}
\end{equation}
where $J_n(x)$ is a Bessel function of order $n$.
Note that the trace-free term is analytic at $r=0$ due to the small-$r$
behaviour of $J_2(lr)$. Following Ref.~\cite{Seljak:1996ve}, let us denote
$\langle \valpha(\vx) \cdot \valpha(\vx')\rangle$ by $\Cgl(r)$ so that
\begin{equation}
\Cgl(r) = \frac{1}{2\pi} \int \ud l\, l^3 C_l^\psi J_0(lr).
\label{eq:11a}
\end{equation}
Similarly we define the anisotropic coefficient
\begin{equation}
\Cgltwo(r) = \frac{1}{2\pi} \int \ud l\, l^3 C_l^\psi J_2(lr),
\label{Cgltwo_flatdef}
\end{equation}
so that
\begin{equation}
\langle \alpha_i(\vx) \alpha_j(\vx') \rangle = \frac{1}{2}
\Cgl(r) \delta_{ij} - \Cgltwo(r) \hat{r}_{\langle i}
\hat{r}_{j \rangle}.
\end{equation}
The lensed correlation function $\xil(r)$ is given by
\begin{eqnarray}
\xil(r) &\equiv&
\langle \tilde{\Theta}(\vx) \tilde{\Theta}(\vx') \rangle \nonumber \\
&=& \int \frac{\ud^2 \vl}{(2\pi)^2} C_l^\Theta e^{i \vl \cdot \vr}
\langle e^{i\vl \cdot [\valpha(\vx) - \valpha(\vx')]} \rangle,
\label{eq:12}
\end{eqnarray}
where we have assumed that the CMB and lensing potential are independent
(i.e.\ we are neglecting the large scale correlation that arises from
the integrated-Sachs-Wolfe effect and has only a tiny effect on the
lensed CMB). Since we are assuming $\valpha$ is a Gaussian field,
$\vl \cdot [\valpha(\vx) - \valpha(\vx')]$ is a Gaussian variate and the
expectation value in Eq.~(\ref{eq:12}) reduces to
\begin{eqnarray}
\langle e^{i\vl \cdot [\valpha(\vx) - \valpha(\vx')]} \rangle &=&
\exp\left(- \frac{1}{2} \langle [\vl \cdot (\valpha-\valpha')]^2 \rangle
\right) \nonumber \\
&=& \exp\left(-\frac{1}{2}l^2 [\sigma^2(r)
+ \cos 2(\phi_\vl -\phi_\vr) \Cgltwo(r)]\right),
\label{eq:13}
\end{eqnarray}
where we have used $l^i l^j \hat{r}_{\langle i} \hat{r}_{j \rangle} =
l^2 \cos 2(\phi_\vl - \phi_\vr) /2$ and defined  $\sigma^2(r) \equiv
\Cgl(0)-\Cgl(r)$. Here, e.g.\ $\phi_\vl$ is the angle
between $\vl$ and the $x$-axis.  The $\cos
2(\phi_\vl -\phi_\vr)$ term in Eq.~(\ref{eq:13}) is difficult to handle
analytically. Instead, we expand the exponential and integrate term by term.
Expanding to second order in $\Cgltwo$, we find
\begin{eqnarray}
\xil(r) &=& \frac{1}{2\pi} \int l \ud l\, C_l e^{-l^2 \sigma^2(r) /2}
\left[ \left(1+\frac{1}{16} l^4 \Cgltwo^2(r)\right)J_0(lr)
+ \frac{1}{2}l^2 \Cgltwo(r) J_2(lr) + \frac{1}{16} l^4 \Cgltwo^2
(r) J_4(lr)\right].
\label{flatt}
\end{eqnarray}
Expanding to this order is sufficient to get the lensed power spectrum to
second order in $C_l^\psi$; higher order terms in $\Cgltwo$ only contribute at the
$O(10^{-4})$ level on the scales of interest.
Note that the $\exp(-l^2 \sigma^2/2)$ term
is easily handled without resorting to a perturbative expansion in
$C_l^\psi$. Since $\sigma^2$ is significantly less than $\Cgltwo$ (as
shown in Fig.~\ref{sigmaplot}), the perturbative expansion in
$\Cgltwo$ converges much faster than one in $\sigma^2$.
Equation~(\ref{flatt}) extends the result of Ref.~\cite{Seljak:1996ve}
to second order in $\Cgltwo$.

\subsubsection{Polarization}

The polarization calculation is also straightforward in the flat-sky
limit~\cite{zaldarriaga98}. We use the spin $-2$ polarization $P \equiv Q +
iU$, where $Q$ and $U$ are the Stokes' parameters measured with
respect to the fixed basis composed of the $x$ and $-y$ axes. Expanding
$P(\vx)$ in terms of the Fourier transforms of its electric ($E$) and
magnetic ($B$) parts, we have
\begin{equation}
P(\vx) = -\int \frac{\ud^2 \vl}{2\pi}\,(E(\vl) - iB(\vl))  e^{-2i\phi_{\vl}}
e^{i\vl\cdot \vx},
\end{equation}
where $(\partial_x-i\partial_y)^2 e^{i\vl\cdot \vx}/l^2= -e^{-2i\phi_{\vl}}
e^{i\vl\cdot \vx}$ is a spin -2 flat-sky harmonic.
The polarization correlation functions are defined as
\begin{eqnarray}
\xi_+(r) &\equiv& \langle e^{-2i\phi_\vr} P^*(\vx) e^{2i \phi_\vr} P(\vx')
\rangle, \label{eq:14b} \\
\xi_-(r) &\equiv& \langle e^{2i \phi_\vr} P(\vx) e^{2i \phi_\vr} P(\vx')
\rangle , \label{eq:14c} \\
\xi_X(r) &\equiv& \langle \Theta(\vx) e^{2i \phi_\vr} P(\vx') \rangle,
\label{flatpol}
\end{eqnarray}
where $\pi - \phi_\vr$ is the angle to rotate the $x$-axis onto the vector
joining $\vx$ and $\vx'$, so that e.g.\ $e^{2i \phi_\vr} P(\vx)$ is the
polarization on the basis adapted to $\vx$ and $\vx'$.
Then the lensed correlation functions to second-order in $\Cgltwo$ are
\begin{eqnarray}
\xil_+(r) &=& \frac{1}{2\pi} \int l \ud l\, (C_l^E + C_l^B)
 e^{-l^2 \sigma^2(r) /2}
\left[ \left(1+\frac{1}{16} l^4 \Cgltwo^2(r)\right)J_0(lr)\right.\nonumber\\
&&\phantom{ \frac{1}{2\pi} \int l \ud l\, (C_l^E + C_l^B)e^{-l^2 \sigma^2(r) /2}+}\left.
+ \frac{1}{2}l^2 \Cgltwo(r) J_2(lr)  + \frac{1}{16} l^4 \Cgltwo^2
(r) J_4(lr)\right], \label{eq:14e} \\
\xil_-(r) &=& \frac{1}{2\pi} \int l \ud l\, (C_l^E - C_l^B)
 e^{-l^2 \sigma^2(r) /2}
\left[ \left(1+\frac{1}{16} l^4 \Cgltwo^2(r)\right)J_4(lr)
\right. \nonumber \\
&&\phantom{ \frac{1}{2\pi} \int l \ud l\, (C_l^E - C_l^B)e^{-l^2 \sigma^2(r) /2}+}\left.
+ \frac{1}{2}l^2 \Cgltwo(r) \frac{1}{2}[J_2(lr)+J_6(lr)]
+ \frac{1}{16} l^4 \Cgltwo^2
(r) \frac{1}{2}[J_0(lr)+J_8(lr)]\right] , \label{eq:14f} \\
\xil_X(r) &=& \frac{1}{2\pi} \int l \ud l\, C_l^{X}
e^{-l^2 \sigma^2(r) /2}
\left[ \left(1+\frac{1}{16} l^4 \Cgltwo^2(r)\right)J_2(lr)
+ \frac{1}{2}l^2 \Cgltwo(r) \frac{1}{2}[J_0(lr)+J_4(lr)] \right.
\nonumber \\
&&\phantom{\frac{1}{2\pi} \int l \ud l\, C_l^{X}
e^{-l^2 \sigma^2(r) /2}+} \left. + \frac{1}{16} l^4 \Cgltwo^2
(r) \frac{1}{2}[J_2(lr)+J_6(lr)]\right] . \label{eq:14g}
\end{eqnarray}
Here $C_l^E$ and $C_l^B$ are the $E$-mode and $B$-mode power spectra, and
$C_l^X$ is the $\Theta$-$E$ cross-correlation. This is the straightforward
extension of the result in Ref.~\cite{zaldarriaga98} to higher
order;\footnote{Note that we disagree with the statement in
Ref.~\cite{zaldarriaga98} that a $O(C_l^\psi)$ expansion is very accurate.
Indeed \CMBFAST\ 4.5 actually uses the non-perturbative $\sigma^2$ term
(as advocated here) rather than the lowest-order series expansion given
in the paper.} see that paper for further details of the calculation.

The lensed $\xil_+(r)$ has the same structure as for the temperature since
the unlensed correlation functions involve the same $J_0(lr)$, and there are
no complications due to the different local bases defined by the
displacement $\vr$ and its image under lensing $\vr - \valpha' + \valpha$
since the phase factors from the rotations cancel.
This is not the case for the lensed
$\tilde{\xi}_-(r)$ and $\tilde{\xi}_X(r)$.

\subsubsection{Limber approximation}

At high $l$ the power spectrum $\clp_\Psi(k)$ varies slowly compared to the
spherical Bessel functions in Eq.~(\ref{cpsi}), which pick out the scale
$k \sim l/\chi$. Using
\begin{equation}
\int k^2 \ud k \, j_l(k\chi) j_l(k\chi') = \frac{\pi}{2\chi^2}
\delta(\chi-\chi') ,
\end{equation}
we can Limber-approximate $C_l^\psi$ as
\begin{equation}
C_l^\psi \approx \frac{8\pi^2}{l^3} \int_0^{\chi_*} \chi \ud \chi\,
\clp_\Psi(l/\chi;\eta_0-\chi) \left(\frac{\chi_*-\chi}{\chi_*\chi}\right)^2.
\end{equation}
Changing variables to $k = l/\chi$, we find
\begin{equation}
\Cgl(r) \approx 4\pi \int \ud k\, \int \ud \chi \clp_\Psi(k;\eta_0-\chi)
\left(\frac{\chi_*-\chi}{\chi_*}\right)^2 J_0(k\chi r),
\label{eq:11b}
\end{equation}
in agreement with Ref.~\cite{Seljak:1996ve} if we note that his $P_\phi(k) =
\clp_\Psi(k)/(4\pi k^3)$ outside radiation-domination.
Ref.~\cite{Seljak:1996ve} also defines $\Cgltwo(r)$ as (in our notation)
\begin{equation}
\Cgltwo(r) \approx 4\pi \int \ud k\, \int \ud \chi \clp_\Psi(k;\eta_0-\chi)
\left(\frac{\chi_*-\chi}{\chi_*}\right)^2 J_2(k\chi r) ,
\label{eq:12b}
\end{equation}
which is the Limber-approximation version of
Eq.~\eqref{Cgltwo_flatdef}. For the results of this paper we do not
use the Limber approximation, though the approximation is rather good.

\subsection{Spherical sky}
\label{spherical}

The flat-sky result for the lensed correlation function is non-perturbative
in $\sigma^2(r)$ and this turns out to be crucial for getting high accuracy
in the lensed power spectrum on arcminute scales. Consider the
contribution to the lensed correlation functions from the unlensed CMB at
multipole $l$. Both $\sigma^2$ and $\Cgltwo$ appear with a factor
$l^2$ and so the (dominant) $l^2 \sigma^2$ term cannot be
handled accurately with a low-order expansion at high $l$.
Physically, this is because the typical lensing displacement is then
comparable to the wavelength of the unlensed fluctuation, and so
approximating the fluctuation as a gradient over the scale of the lensing
displacement is inaccurate. The error from this gradient approximation
on the lensed power spectra will be large on any scale $|\vl |$ where the
dominant contribution is from unlensed fluctuations with wavenumber
$\vl'$ comparable to the typical lensing displacement at scale $|\vl - \vl'|$.

As noted in the Introduction, the small-scale cut-off in the power in the
unlensed fluctuations due to diffusion damping means that the gradient
approximation should not get uniformly worse on small scales; the approximation
should be poorest on scales of a few arcminutes. We also noted that the
the flat-sky approximation will be suspect on large scales, and also
on any scale where the dominant contribution is from large-scale lenses,
i.e.\  those for which their mode-coupled wavenumber $|\vl - \vl'|$ small.
What is needed for an accurate calculation (on all scales where
non-linearities in the lensing potential are not important), is a
non-perturbative treatment of $\sigma^2$ and a proper treatment of
curvature effects in the correlation functions. In this section
we show how to generalize the flat-sky calculation to spherical
correlation functions.

On the full sky we can expand the temperature field in spherical
harmonics
\begin{equation}
\Theta(\vnhat) = \sum_{lm} \Theta_{lm} Y_{lm}(\vnhat) ,
\end{equation}
and the temperature correlation function is defined by
\begin{equation}
\xi(\beta) \equiv \la \Theta(\vnhat_1) \Theta(\vnhat_2) \ra ,
\end{equation}
where $\beta$ is the angle between the two directions ($\vnhat_1 \cdot
\vnhat_2 = \cos\beta$). The power spectrum is defined as the variance
of the harmonic coefficients $C_l^\Theta \equiv \la
|\Theta_{lm}|^2\ra$ for a statistically-isotropic ensemble.

\begin{figure}
\begin{center}
\psfig{figure=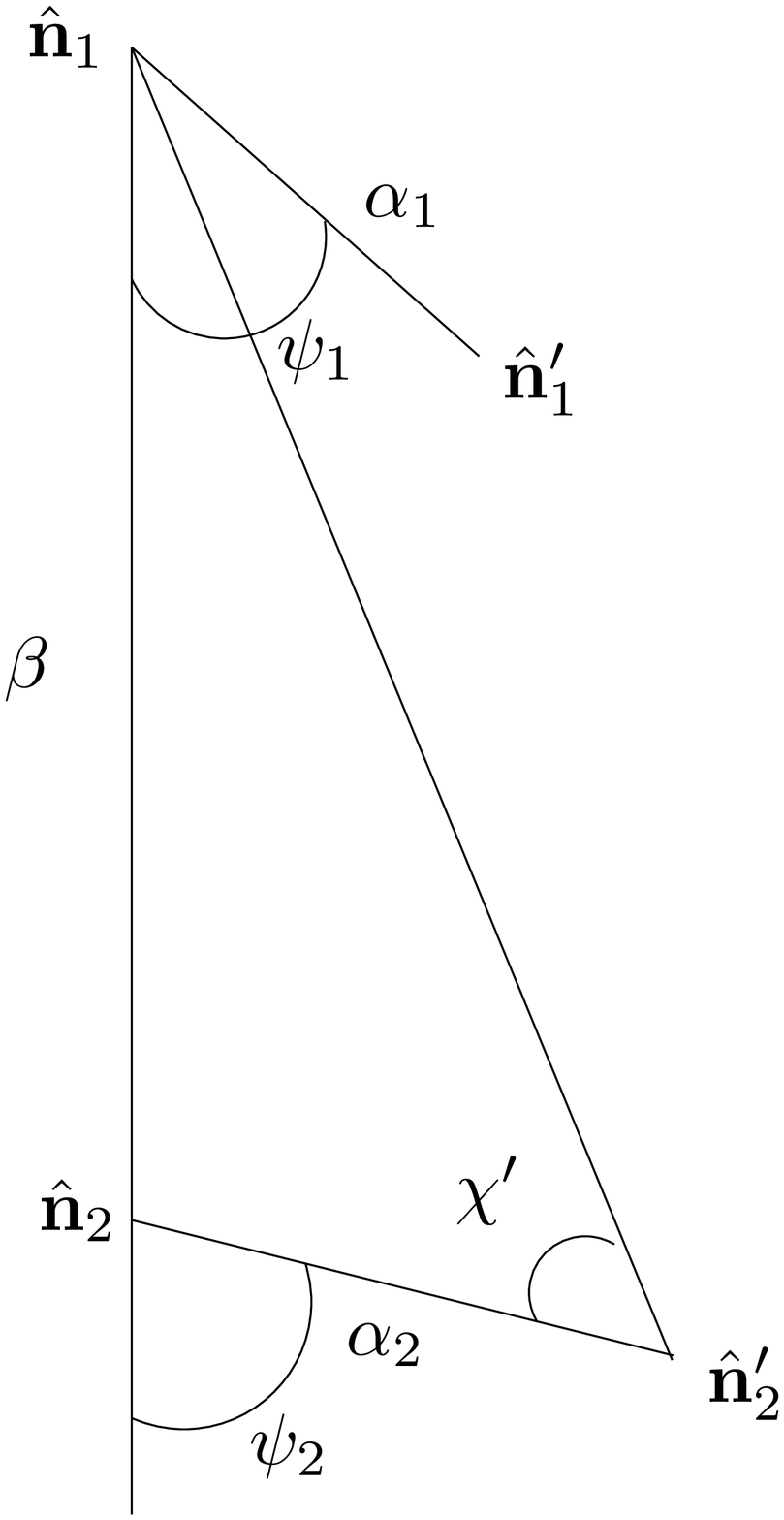,angle=0,width = 4cm}
\caption{The geometry of the weak lensing deflections (shown without
curvature for clarity).
\label{geom}}
\end{center}
\end{figure}

We define a spin-1 deflection field ${}_1 \alpha\equiv \valpha \cdot
(\ve_\theta + i \ve_\phi)$, where $\ve_\theta$ and $\ve_\phi$
are the unit basis vectors of a spherical-polar coordinate system.
Rotating to the basis defined by the geodesic connecting $\vnhat_1$ and
$\vnhat_2$, the spin-1 deflection (denoted with an overbar in the geodesic
basis) has real and imaginary components
\begin{equation}
\alpha_1 \cos\psi_1 = \Re {}_1 \bar{\alpha}(\vnhat_1), \quad
\alpha_1 \sin\psi_1 = \Im {}_1 \bar{\alpha}(\vnhat_1), \quad
\end{equation}
and similarly at $\vnhat_2$. Here,
$\alpha_1 = |\valpha(\vnhat_1)|$ is the length of the lensing displacement
at $\vnhat_1$ and $\psi_1$ is the angle it makes with the geodesic from
$\vnhat_1$ to $\vnhat_2$ (see Fig.~\ref{geom}).
In terms of these angles we have the lensed correlation function
\begin{eqnarray}
\xil(\beta) &=& \la \Theta(\vnhat_1') \Theta(\vnhat_2') \ra\\& =&
 \sum_{lm} C_l^\Theta \langle Y_{lm}
(\vnhat_1') Y^*_{lm}(\vnhat_2') \rangle\\&=&
\sum_{lmm'} C_l^\Theta d^l_{mm'}(\beta) \langle Y_{lm}
(\alpha_1,\psi_1) Y^*_{lm'}(\alpha_2,\psi_2) \rangle.
\label{eq:21}
\end{eqnarray}
The easiest way to see the last step is to put $\vnhat_1$ along the $z$-axis, and
$\vnhat_2$ in the $x$-$z$ plane so that $\vnhat_1'$ has polar coordinates
$(\alpha_1,\psi_1)$. The harmonic at the deflected position
$\vnhat_2'$ can be
evaluated by rotation: $Y_{lm}(\vnhat_2') = [\hat{D}^{-1}(0,\beta,0)
Y_{lm}](\alpha_2,\psi_2)$, where $[\hat{D}Y_{lm}](\vnhat)$ is a spherical
harmonic rotated by the indicated Euler angles.
We have
neglected the small correlation between the deflection angle and
the temperature so that they may be treated as independent fields. The remaining average is over possible realizations of the lensing field.

We assume the lensing potential is Gaussian, so the covariance of the
spin-1 deflection field can be determined using the results
\begin{eqnarray}
\langle {}_1 \bar{\alpha}(\vnhat_1) {}_1 \bar{\alpha}(\vnhat_2) \rangle &=&
- \sum_l \frac{2l+1}{4\pi} l(l+1) C_l^\psi d^l_{-1 1}(\beta) \equiv -
\Cgltwo(\beta) , \nonumber \\
\langle {}_1 \bar{\alpha}^*(\vnhat_1) {}_1 \bar{\alpha}(\vnhat_2) \rangle &=&
\sum_l \frac{2l+1}{4\pi} l(l+1) C_l^\psi d^l_{1 1}(\beta) \equiv
\Cgl(\beta). \label{cgl_def}
\end{eqnarray}
As in the flat-sky limit, it is convenient to define $\sigma^2(\beta)
\equiv \Cgl(0) - \Cgl(\beta)$. The covariance of the Gaussian
variates $\Re {}_1 \bar{\alpha}(\vnhat_1)$,
$\Im {}_1 \bar{\alpha}(\vnhat_1)$, $\Re {}_1 \bar{\alpha}(\vnhat_2)$
and $\Im {}_1 \bar{\alpha}(\vnhat_2)$ are determined by
Eq.~(\ref{cgl_def}). Transforming variables to $\alpha_1$, $\psi_1$,
$\alpha_2$ and $\psi_2$ we find their probability distribution function
\begin{eqnarray}
\text{Pr}(\alpha_1,\alpha_2,\psi_1,\psi_2) &=&
\frac{4 \alpha_1 \alpha_2}{(2\pi)^2}
\frac{e^{-\half(\alpha_1\cos\psi_1
+\alpha_2\cos\psi_2)^2/(\sigma^2+2 \Cgl - \Cgltwo)}}
{\sqrt{\sigma^2+2 \Cgl - \Cgltwo}} \nonumber \\
&&\mbox{}\times
\frac{e^{-\half(\alpha_1\sin\psi_1
+\alpha_2\sin\psi_2)^2/(\sigma^2+2 \Cgl + \Cgltwo)}}
{\sqrt{\sigma^2+2 \Cgl + \Cgltwo}}
\frac{e^{-\half(\alpha_1\cos\psi_1
-\alpha_2\cos\psi_2)^2/(\sigma^2+\Cgltwo)}}
{\sqrt{\sigma^2+\Cgltwo}} \nonumber \\
&&\mbox{}\times
\frac{e^{-\half(\alpha_1\sin\psi_1
-\alpha_2\sin\psi_2)^2/(\sigma^2-\Cgltwo)}}
{\sqrt{\sigma^2-\Cgltwo}} .
\label{eq:22}
\end{eqnarray}
Here and below we have left the dependence of $\sigma^2$, $\Cgl$ and $\Cgltwo$
on $\beta$ implicit.
Our general strategy to evaluate Eq.~(\ref{eq:21}) is to
expand $\text{Pr}(\alpha_1,\alpha_2,\psi_1,\psi_2)$
in $\Cgl$ and $\Cgltwo$, but not $\sigma^2$, before performing the
integral over the angles $\psi_1$ and $\psi_2$ in the expectation value.
The remaining integrals over $\alpha_1$ and $\alpha_2$ then enter through
functions of the form
\begin{equation}
X_{imn} \equiv \int_0^\infty \frac{2\alpha}{\sigma^2}\left(\frac{\alpha}
{\sigma^2}\right)^i e^{-\alpha^2/\sigma^2} d^l_{mn}(\alpha) \, \ud \alpha.
\label{eq:37}
\end{equation}

Since terms
involving $\Cgl(\beta)$ are suppressed at high $l$ (they do not appear in
the flat-sky results), while at low $l$ the leading-order result neglecting
$\Cgl$ and $\Cgltwo$ altogether is very accurate, we neglect terms
involving $\Cgl$ entirely. This approximation is very accurate
[$< O(10^{-4})$] (for completeness the full second-order result is given in
Appendix~\ref{app:fullres}). As shown in Fig.~\ref{sigmaplot} the values of
$\Cgltwo$ are much smaller than $\sigma^2$, so a perturbative
treatment in $\Cgltwo$ is sufficient as in the flat-sky case.
Working to second-order in $\Cgltwo$, we find
\begin{eqnarray}
\xil \approx \sum_l \frac{2l+1}{4\pi} C^\Theta_l\biggl\{
X_{000}^2 d^l_{00} + \frac{8}{l(l+1)} \Cgltwo X_{000}'^{\,2} d_{1-1}^l +
\Cgltwo^2 \left( X_{000}'^{\,2} d^l_{00}+ X_{220}^2 d_{2-2}^l\right)
\biggr\} ,
\label{curvt}
\end{eqnarray}
where primes denote differentiation with respect to $\sigma^2$ [note that
the $X_{imn}$ are implicit functions of $\beta$ via the
dependence on $\sigma^2(\beta)$].
In Appendix~\ref{appb} we develop approximations for the integrals
$X_{imn}$ which are accurate for all $l$. Applying these approximations,
the required $X_{imn}$ are
\begin{eqnarray}
 X_{000}  &\approx& e^{-l(l+1)\sigma^2/4} \\
 X_{220} &\approx& \frac{1}{4} \sqrt{ (l+2)(l-1)l(l+1) }
 e^{-[l(l+1)-2]\sigma^2/4}.
\end{eqnarray}
The expansion of these results to $O(\sigma^2)$ may also be derived
straightforwardly by using the series expansion of
$d^l_{mn}(\alpha)$ for small $\alpha$. (The
smallness of $\sigma^2$ guarantees that the
integral is dominated by the small $\alpha$ region). However, it is
important to retain the correct non-perturbative form for high $l$.

In the limit of large $l$ the limiting result $d^l_{mn}(\beta)
\rightarrow (-1)^{n-m} J_{m-n}(l\beta)$ shows that the full
result of Eq.~\eqref{curvt} reduces to Eq.~\eqref{flatt} in the flat-sky
limit and is therefore consistent. In the limit in which the
separation angle $\beta \rightarrow 0$ we have
\begin{eqnarray}
\xil(0) &=& \sum_l \frac{2l+1}{4\pi} C^\Theta_l =
\sum_l \frac{2l+1}{4\pi} \tilde{C}^\Theta_l ,
\end{eqnarray}
where $\tilde{C}_l^\Theta$ is the lensed power spectrum.
This expresses the fact that weak lensing does not change the
total fluctuation power.

\subsubsection{Polarization}

We can extend the previous calculation to polarization. Defining
Stokes' parameters with the local $x$-axis along the $\theta$-direction
and $y$ along $-\phi$, the quantities $Q \pm i U$ are spin $\mp 2$
respectively. We can expand $Q \pm i U$ in terms of the
spin-weight harmonics as~\cite{Zaldarriaga97}
\begin{equation}
(Q \pm i U )(\vnhat) = \sum_{lm} (E_{lm} \mp i B_{lm}) {}_{\mp 2}
Y_{lm}(\vnhat) ,
\label{eq:qiu}
\end{equation}
which expresses $P \equiv Q + i U$ as the sum of its electric
($E$ or gradient-like) and magnetic ($B$ or curl-like) parts. (Our
conventions for the polarization harmonics and correlation functions
follow Refs.~\cite{Lewis01,Chon:2003gx}; see these papers for a more
thorough introduction).
The polarization correlation functions
can be defined in terms of the spin $\pm 2$ polarization defined in
the physical basis of the geodesic connecting the two positions.
As for the temperature, we evaluate the polarization correlation functions by
taking $\vnhat_1$ along the $z$-axis and $\vnhat_2$ in the $x$-$z$ plane at
angle $\beta$ to the $z$-axis.
With this geometry, the polar-coordinate basis is
already the geodesic basis connecting $\vnhat_1$ and $\vnhat_2$ so
that the lensed correlation functions are
\begin{eqnarray}
\xil_+(\beta) &\equiv & \langle \tilde{P}^*(\vnhat_1)
\tilde{P}(\vnhat_2) \rangle ,\label{eq:27} \\
\xil_-(\beta) &\equiv& \langle \tilde{P}(\vnhat_1)
\tilde{P}(\vnhat_2) \rangle ,\label{eq:28} \\
\xil_X(\beta) &\equiv& \langle \tilde{\Theta}(\vnhat_1)
\tilde{P}(\vnhat_2)\rangle . \label{eq:29}
\end{eqnarray}
Under a lensing deflection the polarization orientation is preserved
relative to the direction of the deflection (we are neglecting the
small effect of field rotation~\cite{Hirata:2003ka}), i.e. the
polarization undergoes parallel transport. The geometry of the
deflections is shown in Fig.~\ref{geom}.

We can easily evaluate the lensed polarization on
the connecting geodesic basis (between $\vnhat_1$ and $\vnhat_2$)
as
\begin{equation}
\tilde{P}(\vnhat_1) = P(\alpha_1,\psi_1) e^{-2i\psi_1}.
\end{equation}
The rotation
angle $\psi_1$ is that needed to rotate the spin $-2$ polarization from polar
coordinates (coinciding with the the $\vnhat_1$--$\vnhat_1'$ basis at
$\vnhat_1'$) to the geodesic basis connecting $\vnhat_1$ and $\vnhat_2$.

For the lensed polarization at $\vnhat_2$ a little more work is required.
Let $\chi'$ denote the angle between the geodesics connecting $\vnhat_2$
to $\vnhat_2'$, and $\vnhat_1$ (along the $z$-axis) to
$\vnhat_2'$ (see Fig.~\ref{geom}).
The lensed polarization at $\vnhat_2$ on the geodesic basis adapted to $\vnhat_1$ and
$\vnhat_2$ is then
\begin{equation}
\tilde{P}(\vnhat_2) = P(\vnhat_2') e^{2i\chi'} e^{-2i\psi_2}.
\label{eq:30}
\end{equation}
We can write $\vnhat_2'$ as the direction obtained
by rotating a direction with polar angles $(\alpha_2,\psi_2)$ by
an angle $\beta$ about
the $y$-axis, i.e.\ $\vnhat_2' = \hat{D}(0,\beta,0)(\alpha_2,\psi_2)$.
Writing $P$ as $(Q-iU)^*$, and using Eq.~(\ref{eq:qiu}), we have
\begin{equation}
P(\vnhat) = \sum_{lm} (E_{lm} + i B_{lm})^* {}_{+2} Y_{lm}^*(\vnhat).
\label{eq:31}
\end{equation}
Using the rotation properties of the spin-$s$ harmonics (see
Appendix~\ref{appa}), we then find
\begin{equation}
\tilde{P}(\vnhat_2) = e^{2i\chi'} e^{-2i\psi_2} e^{-2i\kappa} \sum_{lmm'}
(E_{lm} + i B_{lm})^* D^{l*}_{mm'}(0,\beta,0) {}_2 Y_{lm'}^*(\alpha_2,\psi_2).
\label{eq:32}
\end{equation}
The angle $\kappa$ is the rotation about $\vnhat_2'$ that is required to
bring the polar basis there onto that obtained by rotating the polar basis
at $(\alpha_2,\psi_2)$ with $\hat{D}(0,\beta,0)$. Since the latter is aligned
with the geodesic basis adapted to $\vnhat_2$ and $\vnhat_2'$, we have
$\kappa=\chi'$ and the lensed polarization at $\vnhat_2$ simplifies to
\begin{equation}
\tilde{P}(\vnhat_2) =  e^{-2i\psi_2} \sum_{lmm'}
(E_{lm} + i B_{lm})^* d^l_{mm'}(\beta) {}_2 Y_{lm'}^*(\alpha_2,\psi_2).
\label{eq:33}
\end{equation}
We can now quickly proceed to the following expressions for the lensed
polarization correlation functions:
\begin{eqnarray}
\xil_+(\beta) &=& \sum_{lmm'} (C_l^E+C_l^B) d^l_{mm'}(\beta)
\langle e^{2i\psi_1} {}_2 Y_{lm}(\alpha_1,\psi_1) {}_2 Y_{lm'}^*
(\alpha_2,\psi_2) e^{-2i\psi_2} \rangle , \label{eq:34} \\
\xil_-(\beta) &=& \sum_{lmm'} (C_l^E-C_l^B) d^l_{mm'}(\beta)
\langle e^{-2i\psi_1} {}_{-2} Y_{lm}(\alpha_1,\psi_1) {}_2 Y_{lm'}^*
(\alpha_2,\psi_2) e^{-2i\psi_2} \rangle , \label{eq:35} \\
\xil_X(\beta) &=& \sum_{lmm'} C_l^X d^l_{mm'}(\beta)
\langle Y_{lm}(\alpha_1,\psi_1) {}_2 Y_{lm'}^*
(\alpha_2,\psi_2) e^{-2i\psi_2} \rangle , \label{eq:36}
\end{eqnarray}
where the expectation values are over lensing realizations. Here,
$C_l^E$ and $C_l^B$ are the power spectra $\langle |E_{lm}|^2 \rangle$
and $\langle |B_{lm}|^2 \rangle$ respectively. The cross-correlation
power spectrum is $C_l^X \equiv \langle \Theta_{lm} E_{lm}^* \rangle$.

We evaluate the expectation values in Eqs.~(\ref{eq:34}--\ref{eq:36})
following the earlier calculation for the temperature, i.e.\ expanding
$\text{Pr}(\alpha_1,\alpha_2,\psi_1,\psi_2)$ to second order in
$\Cgltwo$ before integrating. As for the temperature, $\Cgl$ terms
contribute negligibly (see Appendix~\ref{app:fullres} for the full result).
We find the following results for the lensed polarization correlation
functions to second order in $\Cgltwo$:
\begin{eqnarray}
\xil_+\angarg &\approx& \sum_{lmm'} \frac{2l+1}{4\pi}
(C_l^E + C_l^B)\Big\{ X_{022}^2 d^l_{22}\angarg +2 \Cgltwo\angarg
X_{132}X_{121} d^l_{31}\angarg
+ \Cgltwo^2\angarg[(X_{022}')^2 d^l_{22}
\angarg + X_{242}X_{220} d^l_{40}\angarg] \Big\} , \label{eq:38} \\
\xil_-\angarg &\approx& \sum_{lmm'} \frac{2l+1}{4\pi}
(C_l^E - C_l^B)\Bigg\{ X_{022}^2 d^l_{2\, -2}\angarg +
\Cgltwo\angarg[X_{121}^2 d^l_{1\,-1}\angarg + X_{132}^2 d^l_{3\,-3}\angarg]
\nonumber \\
&&\phantom{ \sum_{lmm'} \frac{2l+1}{4\pi}(C_l^E -
  C_l^B)\Bigg\{}\mbox{}
+ \frac{1}{2}\Cgltwo^2[2 (X_{022}')^2 d^l_{2\,-2}\angarg +
X_{220}^2 d^l_{00}\angarg + X_{242}^2 d^l_{4\,-4}\angarg] \Bigg\} ,
\label{eq:39} \\
\xil_X\angarg &\approx& \sum_{lmm'} \frac{2l+1}{4\pi}
C_l^X \Bigg\{X_{022}X_{000} d^l_{02}\angarg + \Cgltwo\angarg\frac{2 X_{000}'}
{\sqrt{l(l+1)}} (X_{112} d^l_{11}\angarg + X_{132}d^l_{3\,-1}\angarg )
\nonumber \\
&&\phantom{ \sum_{lmm'} \frac{2l+1}{4\pi}C_l^X \Bigg\{}\mbox{}
 + \frac{1}{2} \Cgltwo^2\angarg[(2X_{022}'X_{000}'+X_{220}^2)
d^l_{20}\angarg
+ X_{220} X_{242} d^l_{-2 4}\angarg]\Bigg\},  \label{eq:40}
\end{eqnarray}
where
\begin{eqnarray}
X_{022} & \approx & e^{-[l(l+1)-4]\sigma^2/4} , \\
X_{121} & \approx & -\frac{1}{2} \sqrt{(l+2)(l-1)} e^{-[l(l+1)-8/3]\sigma^2/4}
,  \\
X_{132} & \approx & -\frac{1}{2} \sqrt{(l+3)(l-2)} e^{-[l(l+1)-20/3]\sigma^2/4}
, \label{eq:43} \\
X_{242} & \approx & \frac{1}{4} \sqrt{(l+4)(l+3)(l-2)(l-3)}
e^{-[l(l+1)-10]\sigma^2/4} .
\end{eqnarray}
These expressions for the $X_{imn}$ are accurate to $O(\sigma^2)$ at
low $l$, and
have the correct non-perturbative form at high $l$. Since only $X_{000}$
and $X_{022}$ enter at lowest order, the other exponentials may be
further safely approximated as $\sim X_{000}$ since their contributions
will be negligible at low $l$.

In the limit of zero separation $\beta\rightarrow 0$ we have
\begin{eqnarray}
\xil_+(0) &=& \sum_l \frac{2l+1}{4\pi}(C_l^E + C_l^B) = \sum_l
\frac{2l+1}{4\pi}(\tilde{C}_l^E + \tilde{C}_l^B) ,\\
\xil_-(0) &=& \xil_X(0) = 0 ,
\end{eqnarray}
where $\tilde{C}_l^E$ and $\tilde{C}_l^B$ are the lensed $E$- and $B$-mode
power spectra respectively.
This shows that lensing does not change the total polarization power,
though it mixes $E$ and $B$ modes as well as different scales.

\subsubsection{CMB power spectra}

Once the lensed correlation functions have been computed, transforming
to the CMB power spectra is straightforward using
\begin{eqnarray}
\tilde{C}_l^\Theta &=& 2\pi\int_{-1}^1 \xil(\beta)
d^l_{00}(\beta) \ud \cos\beta , \\
\tilde{C}_l^E - \tilde{C}_l^B &=& 2\pi\int_{-1}^1 \xil_-(\beta)
d^l_{2-2}(\beta) \ud \cos\beta , \\
\tilde{C}_l^E + \tilde{C}_l^B &=& 2\pi\int_{-1}^1 \xil_+(\beta)
d^l_{22}(\beta) \ud \cos\beta , \\
\tilde{C}_l^X  &=& 2\pi\int_{-1}^1 \xil_X(\beta)
d^l_{20}(\beta) \ud \cos\beta .
\end{eqnarray}
For a further discussion of correlation functions and the transform to
power spectra see Ref.~\cite{Chon:2003gx}.

\subsubsection{Numerical implementation}

\begin{figure}
\begin{center}
\psfig{figure=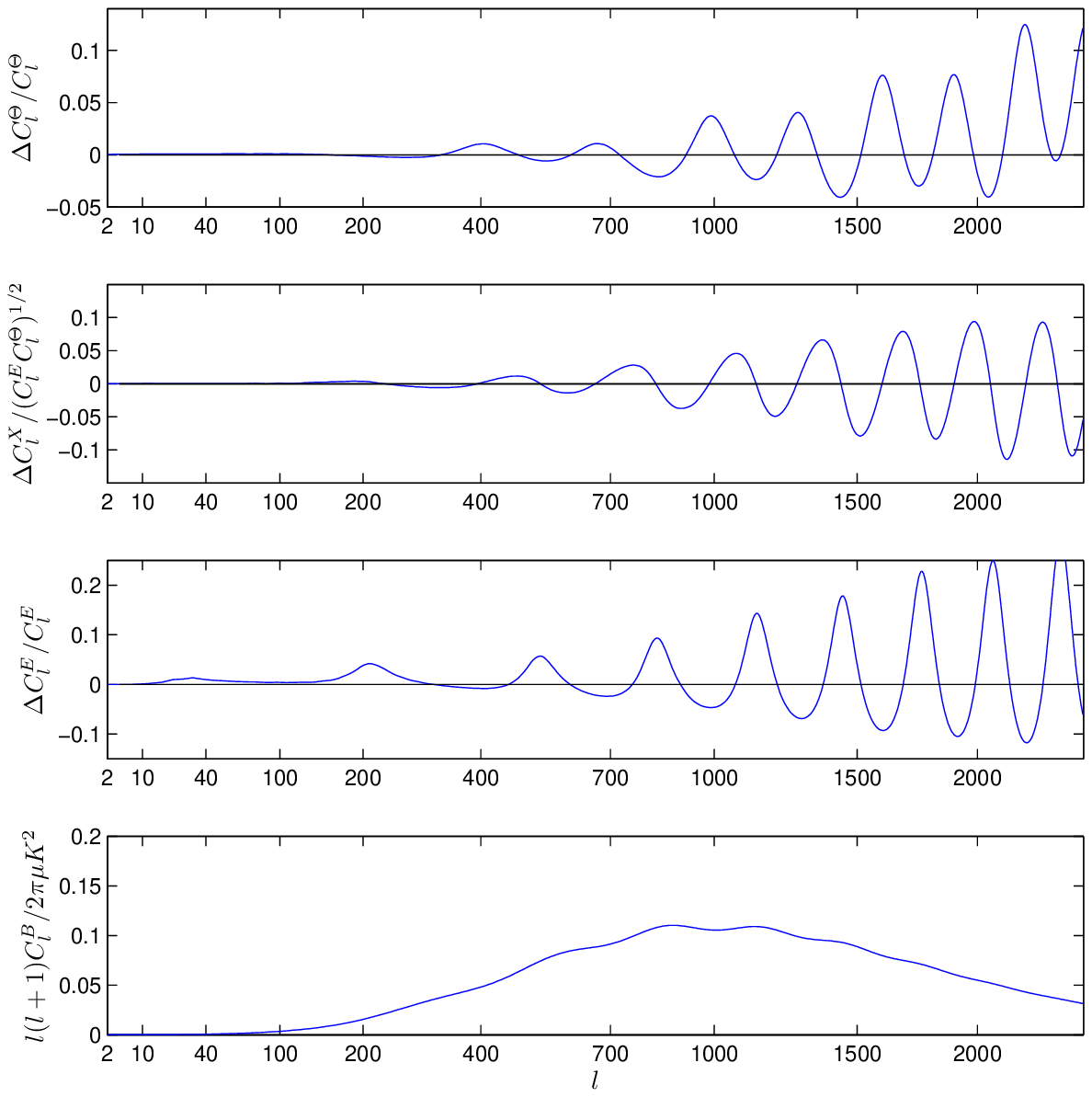,width=13cm}
\caption{Difference between the lensed and unlensed temperature,
cross-correlation and $E$-polarization power spectra (top three plots),
and the lensed $B$ power spectrum (bottom) for a fiducial concordance model.
The unlensed model has no tensor component (so no $B$-mode power), and the
lensed $B$ power spectrum shown is not highly accurate due
to the neglect of non-linear evolution in the lensing potential.
The magnitude of the lensing effect depends on the fluctuation amplitude in
the model; here the model has curvature perturbation power $A_s =
2.5\times 10^{-9}$ on $0.05\,\Mpc^{-1}$ scales and spectral index $n_s=0.99$.
\label{lensedcls}}
\end{center}
\end{figure}

The correlation function method is inherently very efficient, only
requiring the evaluation of one dimensional sums and integrals.
For an accurate calculation of $\tilde{C}_l^B$ it is essential to
compute the full range of the correlation function because it is
sensitive to large and small scales. However, when $\tilde{C}_l^B$ is not
needed the lensing is only a small-scale effect and we only need to
integrate some of the angular range to compute
$\xil(\beta)-\xi(\beta)$ (and hence the lensing contribution
$\tilde{C}_l - C_l$). We find that using
$\beta_{\text{max}}=\pi/16$ is sufficient for $0.1\%$ accuracy to
$l=2000$, providing a significant factor of 16 gain in
speed. Truncating the correlation function does induce ringing on very
small scales, so if accuracy is needed on much smaller scales the
angular range can be increased.
For all but very small scales, and the $\tilde{C}_l^B$ spectrum, we can
accurately evaluate the sums over $l$ to compute the lensed correlation
functions by sampling only every 10th $l$, yielding an
additional significant time saving.

Our code is publicly available as part of
\CAMB,\footnote{\url{http://camb.info}} with execution time being
dominated by that required to compute the transfer functions for the
CMB and the lensing potential. Once these have been computed, the time
required to compute the unlensed $C_l$ and then lens the result is about a
hundred times less (if $\tilde{C}_l^B$ is not required accurately).
This means that efficient methods for exploiting `fast' and `slow'
parameters~\cite{Lewis:2002ah,cosmomc_notes} during
Markov Chain Monte Carlo parameter estimation can still be applied
when lensing is accounted for via the lensed power spectrum.

Sample numerical results for the lensed CMB power spectra compared to
the unlensed spectra are shown in Fig.~\ref{lensedcls}.

\section{Comparison of methods}
\label{sec:comparison}

\begin{figure}
\begin{center}
\psfig{figure=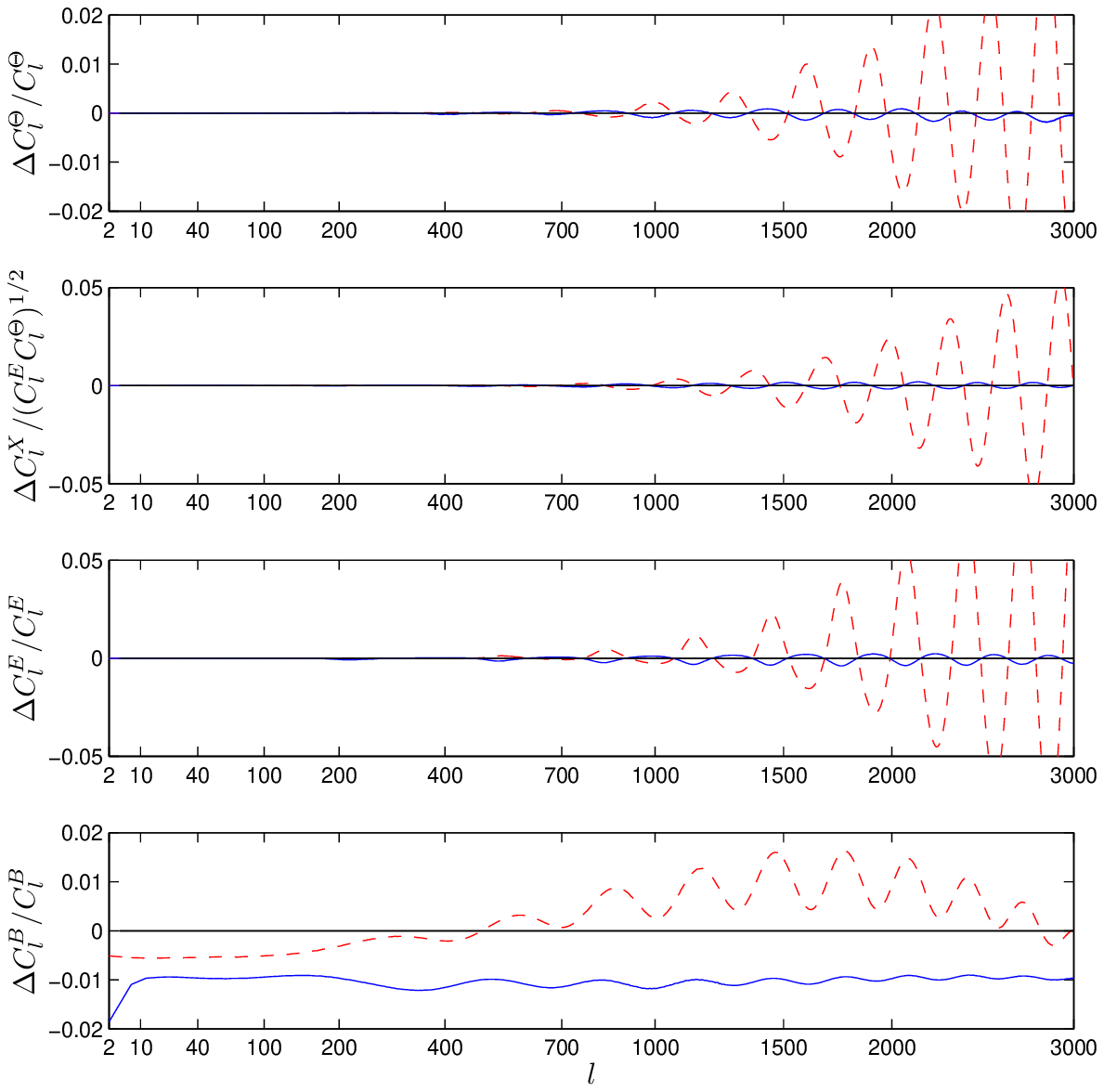,width=13cm}
\caption{
Comparison of our new result with the $O(C_l^\psi)$  harmonic result of
Ref.~\cite{Hu:2000ee} (dashed) and the flat-sky non-perturbative
result of Ref.~\cite{Seljak:1996ve}, extended to second order in
$\Cgltwo$ (solid). The magnitude of the difference depends on the
exact model and we have neglected non-linear contributions to the
lensing potential.
\label{lensedcl_comp}}
\end{center}
\end{figure}

We are now in a position to compare our new, accurate full-sky result
with previous work. In Fig.~\ref{lensedcl_comp} we compare our result
with the full-sky lowest-order perturbative harmonic result of
Ref.~\cite{Hu:2000ee} [correct to $O(C_l^\psi{})$] for a typical concordance
model. We also compare to the flat-sky result of
Refs.~\cite{Seljak:1996ve,Zaldarriaga:1998ar} which is non-perturbative in
$\sigma^2$. [We extend their results to second
order in $\Cgltwo$ using Eqs.~\eqref{flatt} and \eqref{flatpol}].
In all cases we
use an accurate numerical calculation of $C_l^\psi$, rather than
the Limber approximation, and ignore its non-linear contribution.

It is clear that the lowest-order perturbative harmonic method of
Ref.~\cite{Hu:2000ee} is not sufficiently accurate for precision
cosmology, with $\sim 1\%$ errors on the temperature and $\sim 5\%$ on
the $E$-mode polarization by $l\sim 2000$. These errors are sufficient to
bias parameters even with the planned
Planck\footnote{\url{http://sci.esa.int/planck}} satellite
observations. The perturbative harmonic result is equivalent to expanding the
correlation function result self-consistently to first order in
$C_l^\psi$. As discussed in Sec.~\ref{spherical},
this is inaccurate because $l^2\sigma^2$ in the isotropic
terms is not very small for large $l$, so many terms need to be
retained to get accurate results. It is possible to extend the
harmonic result to higher order~\cite{Cooray:2003ar}, however the
multi-dimensional integrals required scale exponentially badly with increasing
order. Even a self-consistent expansion to second order in
$C_l^\psi$ is not good enough at $l > 2000$, so at
least third order would be required. Furthermore we see that for
$\tilde{C}_l^B$ the method is also somewhat inaccurate on large scales:
because the $B$-mode signal comes from a wide range of $l$, and the $E$-mode
power peaks on small scales, the non-perturbative effects can be significant
on all scales. In fact, the large-scale lensed $E$-mode power also receives
most of its contribution from small-scale modes since the unlensed polarization
power spectrum rises steeply with $l$ on large scales. However, lensing
is still only a small fractional effect for $E$-polarization on large scales
and so the perturbative expansion is relatively more accurate for $E$ than
$B$.

The correlation function methods can easily handle the isotropic term
non-perturbatively. The accurate flat-sky result is much more accurate than
the lowest-order harmonic full-sky result, with only $\sim 0.3\%$ curvature
corrections to the temperature.\footnote{Due to the opposite sign of
curvature and second-order corrections in $C_l^\psi$, the flat-sky correlation
result correct to $O(\Cgltwo)$ is actually slightly more accurate than the
result correct to $O(\Cgltwo^2)$.} The polarization errors are rather larger,
with percent-level difference on $\tilde{C}_l^B$. Although this is smaller
than the effect of non-linearities in the lensing potential (see
Section~\ref{sec:nonlin}), the latter can be accurately accounted for
with better modelling (e.g.\ Ref.~\cite{Smith:2002dz}) or simulations.
While the accurate flat-sky result is probably sufficient to
Planck sensitivities, curvature effects must be taken account for truly
accurate results approaching the cosmic-variance limit. Although the curvature
is negligible on the scale of the deflection angles, it is not negligible on
the scale of the lensing potential coherence length.
Computing our full-sky accurate result is not much harder or slower than
computing the flat result, so we recommend our new calculation for future work.

Note that the
absolute precision of the lensed results is limited by the accuracy of the computed lensing potential
and the unlensed  CMB power spectra. In particular,
uncertainties in the ionization history may generate errors
significantly above cosmic variance on the unlensed $C_l$. We use the
\RECFAST\ code of Ref.~\cite{Seager:1999km} that may well be
inaccurate at above the percent
level\footnote{\url{http://cosmocoffee.info/viewtopic.php?t=174}}~\cite{Leung:2003je,Dubrovich:2005fc}.
However if the
ionization history can be computed reliably to high accuracy our new lensing
method can then be used to compute the lensed power spectra accurately.

\section{Non-linear evolution}
\label{sec:nonlin}
\begin{figure}
\begin{center}
\psfig{figure=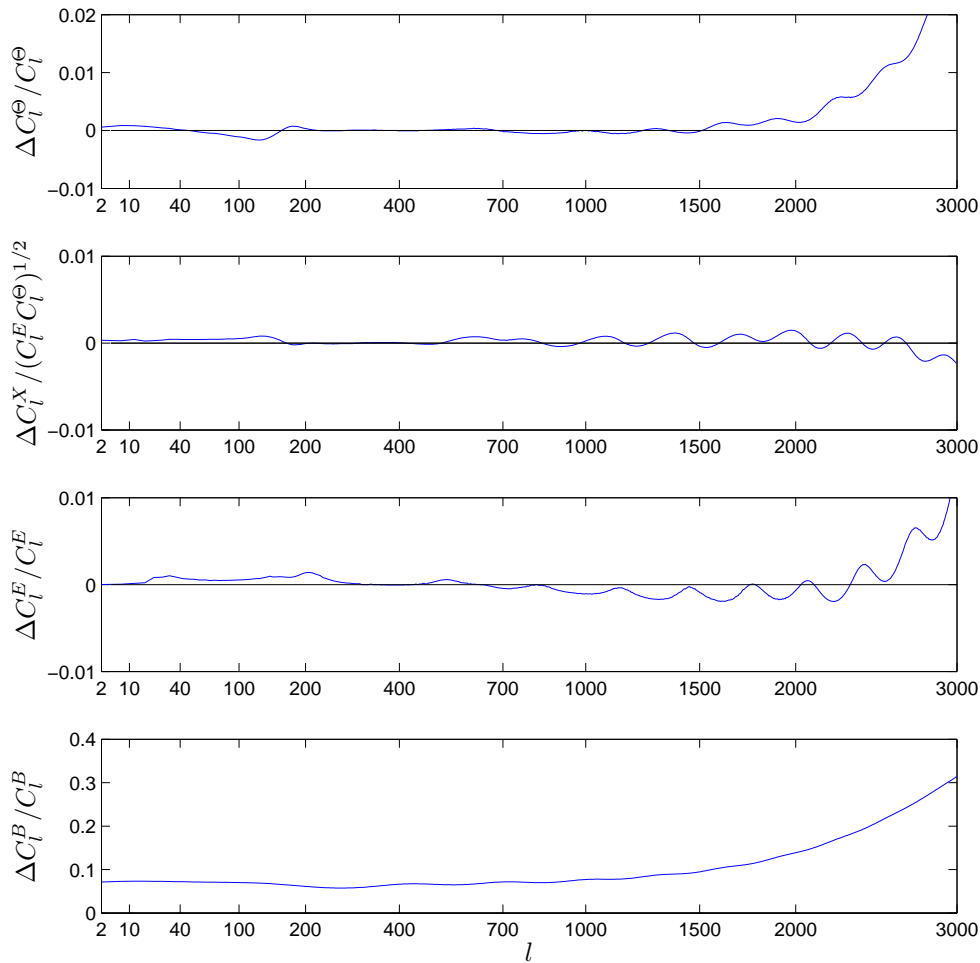,width=13cm}
\caption{The fractional change in the lensed $C_l$ due to non-linear corrections using \HALOFIT~\cite{Smith:2002dz} for the same model as Fig.~\ref{lensedcls}. The lensed $C_l$ are computed using our new accurate method.
\label{nonlin}}
\end{center}
\end{figure}

The most important assumption we have made so far is that the lensing potential is linear and Gaussian. On small scales this will not be quite correct. Although our method does not allow us to account for the non-Gaussianity, we can take into account the effect of non-linear evolution on the power spectrum of the lensing potential [and hence $\sigma^2(\beta)$ and $\Cgltwo(\beta)$~\cite{Seljak:1996ve}].
On scales where the non-Gaussianity of the deflection field is small this should be a good approximation, assuming we have an accurate way to compute the
non-linear power spectrum of the density field.

We use the \HALOFIT\ code of Ref.~\cite{Smith:2002dz} to compute an
approximate non-linear, equal-time power spectrum given an accurate numerical linear power spectrum at a given redshift. \HALOFIT\ is expected to be accurate at the few percent level for standard $\Lambda$CDM models with power law primordial power spectra (but cannot be relied on for other models, for example with an evolving dark energy component).
We simply scale the potential transfer functions $T_\Psi(k,\eta)$ of Eq. \eqref{cpsi_transfer} so that the power spectrum of the potential $\Psi$ has the correct non-linear form at that redshift:
\begin{equation}
T_\Psi(k,\eta) \rightarrow T_\Psi(k,\eta) \sqrt{\frac{\clp^\text{non-linear}_\Psi(k,\eta)}{\clp_\Psi(k,\eta)}}.
\label{Tnonlin}
\end{equation}
Since non-linear effects on $C_l^\psi$ are only important where the Limber
approximation holds, Eq.~(\ref{Tnonlin}) should be very accurate.

The effect of the non-linear evolution on the power spectrum of the lensing potential is shown in Fig.~\ref{CPhi}. Although there is very little effect on scales where the power peaks ($l\sim 60$), non-linear evolution significantly increases the power on small scales. The corresponding changes to the lensed CMB power spectra are shown in Fig.~\ref{nonlin}. The temperature power spectrum $\tilde{C}_l^{\Theta}$ is changed by $\alt 0.2\%$ for $l\sim 2000$, but there are percent level changes on smaller scales.
Thus inclusion of the non-linear evolution will be important to obtain results accurate at cosmic-variance
levels, but is not likely to be important at $l< 2000$ for the near future. The effect on the $B$-mode power spectrum is more dramatic, giving a $>6\%$ increase in power on all scales. On scales beyond the peak in the $B$-mode power
($l\agt 1000$) the extra non-linear power becomes more important, producing an order unity change in the $B$-mode spectrum on small scales. On these scales the assumption of Gaussianity is probably not very good, and the accuracy will also be limited by the precision of the non-linear power spectrum.
For more accurate results, more general models, and on very small scales where the non-Gaussianity of the
lensing potential becomes important, numerical simulations may
be required (e.g. see Refs.~\cite{White:2003xz,White:1999xa}).

There are, of course, other non-linear effects on the CMB with the same
frequency spectrum as the primordial (and lensed) temperature anisotropies
and polarization. The kinematic Sunyaev-Zel'dovich (SZ) effect is the main such
effect for the temperature anisotropies, and current uncertainties in the
reionization history and morphology make the spectrum $C_l^\Theta$ uncertain
at the few percent level at $l=2000$~\cite{Zahn:2005fn}.
This is a little larger than
the error in the first-order harmonic lensing result, but this doesn't mean
that one should be content with the error in the latter. Precision cosmology
from the damping tail will require accurate modelling of both lensing and
the kinematic SZ effect. Errors at the percent level in the lensing power
on these scales would seriously limit our ability to constrain reionization
scenarios with future arcminute-resolution observations. For the polarization
spectra, the kinematic SZ effect is much less significant~\cite{Hu:1999vq}.

\section{Conclusions}

We have presented a new, fast and accurate method for computing the
lensed CMB power spectra using spherical correlation functions. Previous
perturbative methods were found to be insufficiently inaccurate for
precision cosmology, and non-perturbative results in the flat-sky approximation
are in error at above the cosmic-variance level. The method developed here
should enable accurate calculation of the lensing effect to
within cosmic-variance limits to $l \alt 2500$ under the assumptions
of the Born approximation and Gaussianity of the primordial fields.
Non-linear corrections to the lensing potential have only a small effect on
the lensed temperature power spectrum, but are important on all scales
for an accurate calculation of the lensed $B$-mode power spectrum.

\section{Acknowledgments}
We thank Gayoung Chon for her work towards implementing
the full-sky lowest-order lensing result of Ref.~\cite{Hu:2000ee} in
\CAMB, and AL thanks Matias
Zaldarriaga, Mike Nolta,
Oliver Zahn, Patricia Castro, Pat McDonald and Ben Wandelt for discussion and communication.
AC acknowledges a Royal Society University Research Fellowship.

\appendix

\section{Rotating spin-weight harmonics}
\label{appa}

Consider evaluating ${}_s Y_{lm}$ at $\hat{D}\vnhat$, where $\hat{D}$
is the rotation operator corresponding to Euler angles $\alpha$,
$\beta$ and $\gamma$. This is the same as rigidly rotating the function
${}_s Y_{lm}$ (as a scalar) by $\hat{D}^{-1}$ and evaluating at $\vnhat$.
For spin-0 harmonics we know that
\begin{equation}
Y_{lm}(\hat{D}\vnhat) = D^l_{m' m}(-\gamma,-\beta,-\alpha) Y_{lm'}(\vnhat).
\label{eq:app1}
\end{equation}
For spin-$s$ harmonics, we note that
\begin{equation}
{}_s Y_{lm} (\vnhat) = (-1)^m \sqrt{\frac{2l+1}{4\pi}}
D^l_{-m s}(\phi,\theta,0),
\label{eq:app2}
\end{equation}
where $(\theta,\phi)$ refer to $\vnhat$, so that
\begin{eqnarray}
D^l_{m'm}(-\gamma,-\beta,-\alpha) {}_s Y_{lm'}(\vnhat) &=&
(-1)^{m'} \sqrt{\frac{2l+1}{4\pi}} D^l_{m' m}(-\gamma,-\beta,-\alpha)
D^l_{-m' s}(\phi,\theta,0) \nonumber \\
&=& (-1)^{m} \sqrt{\frac{2l+1}{4\pi}} D^l_{-m m'}(\alpha,\beta,\gamma)
D^l_{m' s}(\phi,\theta,0) \nonumber \\
&=& (-1)^{m} \sqrt{\frac{2l+1}{4\pi}} D^l_{-m s}(\phi',\theta',\kappa)
\nonumber \\
&=& {}_s Y_{lm}(\hat{D}\vnhat) e^{-is\kappa}.
\label{eq:app3}
\end{eqnarray}
Here, we have used $\hat{D}(\alpha,\beta,\gamma) \hat{D}(\phi,\theta,0)
= \hat{D}(\phi',\theta',\kappa)$, so that $(\theta',\phi')$ refer to the
image of $\vnhat$ under $\hat{D}(\alpha,\beta,\gamma)$, and $\kappa$ is the
additional rotation required about $\hat{D}\vnhat$ to map the polar basis
vectors there onto the image of the polar basis at $\vnhat$ under
$\hat{D}(\alpha,\beta,\gamma)$. Denoting the polar basis (unit) vectors at
$\vnhat$ by $\ve_\theta$ and $\ve_\phi$, and at
$\vnhat'$ by $\ve_{\theta}'$ and $\ve_{\phi}'$, we have
\begin{eqnarray}
\ve_{\theta}' \pm i \ve_{\phi}' =
e^{\pm i\kappa} \hat{D}(\ve_\theta \pm i \ve_\phi).
\end{eqnarray}
This ensures that the $2l+1$ rank-$s$ tensor fields ${}_{\pm}
\mathbf{Y}_{lm}(\vnhat) \equiv {}_{\pm s} Y_{lm}(\vnhat) (\ve_\theta \mp
i \ve_\phi) \otimes \cdots \otimes (\ve_\theta \mp i \ve_\phi)$
transform irreducibly under rotations as $\hat{D} {}_{\pm} \mathbf{Y}_{lm}
= \sum_{m'} D^l_{m' m} {}_{\pm} \mathbf{Y}_{lm'}$.

\section{Evaluation of $X_{imn}$}
\label{appb}

The integrals
\begin{equation}
X_{imn} \equiv \int_0^\infty \frac{2\alpha}{\sigma^2}\left(\frac{\alpha}
{\sigma^2}\right)^i e^{-\alpha^2/\sigma^2} d^l_{mn}(\alpha) \, \ud \alpha
\label{eq:appb1}
\end{equation}
that are required for the non-perturbative calculation of the lensed
power spectra on the spherical sky can easily be evaluated as series in
$\sigma^2$. From the definition of the rotation matrices, we have
\begin{equation}
d^l_{mn}(\alpha) = \langle l m | e^{-i \alpha \hat{L}_y} | l n \rangle,
\label{eq:appb2}
\end{equation}
where we adopt the Condon--Shortley phase for the eigenstates
$| lm \rangle$ of the $\hat{L}_z$ and $\hat{L}^2$ angular momentum
operators, and we have set $\hbar=1$. Expanding the exponential as a series
in $\alpha$, we have
\begin{equation}
X_{imn} = \sum_{j=0}^\infty \frac{1}{j!} \sigma^{j-i} \Gamma(j + 1 + i/2)
\langle lm | (-i \hat{L}_y)^j | ln \rangle.
\label{eq:appb3}
\end{equation}
The action of the $\hat{L}_y$ operator on the eigenstates of
$\hat{L}_z$ is given by the familiar result
\begin{equation}
-i \hat{L}_y | lm \rangle = \half \sqrt{l(l+1) - m (m-1)}
| l \, m-1 \rangle - \half \sqrt{l(l+1)-m(m+1)} | l \, m+1 \rangle,
\end{equation}
and this can be used recursively to evaluate the matrix element in
Eq.~(\ref{eq:appb3}).

For $l \gg 1/\sigma$ the series in Eq.~(\ref{eq:appb3}) is slow to
converge and a non-perturbative treatment of the $X_{imn}$ is required.
Using the asymptotic result
\begin{equation}
d^l_{mn}(\alpha) \sim (-1)^{n-m} J_{m-n}[(l+1/2)\alpha],
\end{equation}
valid for $l \rightarrow \infty$ with $l \alpha \gg 1$ but $\alpha \ll 1$,
and, noting that we only require the case $i = m-n $,\footnote{Note that
$X_{imn} = (-1)^{m+n} X_{inm}$ so we can always take $m\geq n$.} we can
use Eq.~(6.6314) from Ref.~\cite{Gradshteyn00} to show that
\begin{equation}
X_{imn} \sim \left(-\frac{(l+1/2)}{2}\right)^i e^{-(l+1/2)^2\sigma^2/4}
\quad \mbox{for $i = m-n$}.
\end{equation}
In practice, we obtain an excellent approximation to $X_{imn}$, valid for
all $l$, by adjusting the $l$-independent term in the exponent of the
asymptotic result, and the prefactor, so that its series expansion agrees
with a direct evaluation of Eq.~(\ref{eq:appb3}) to $O(\sigma^2)$.

\section{Full second order result}
\label{app:fullres}

The full result for the lensed correlation functions accurate to second order
in $\Cgl$ and $\Cgltwo$ is\footnote{Maple code to derive this result is
available at \url{http://camb.info/theory.html}.}
\begin{align}
\xil &\approx \sum_l \frac{2l+1}{4\pi} C^{\Theta}_l\biggl\{
\left(X_{000}^2 + 2\Cgl X_{000} X_{000}' + \Cgl^2(X_{000}'' X_{000} +
2 X_{000}'^{\,2}) +\Cgltwo^2 X_{000}'^{\,2}\right) d_{00}^l
\nonumber\\&\hspace{3cm}+ \frac{8}{l(l+1)} X_{000}' \left( X_{000}'+2
  \Cgl X_{000}''\right) \left(\Cgltwo d_{1-1}^l + \Cgl
  d_{11}^l\right) +\Cgltwo^2  X_{220}^2 d_{2-2}^l
\nonumber\\&\hspace{3cm}
   + \Cgl^2 X_{220}^2 d_{22}^l - 4\Cgl\Cgltwo X_{000}' X_{220} d_{20}^l
\biggr\}
\\
\xil_+ &\approx \sum_l \frac{2l+1}{4\pi} (C^E_l +
C^B_l)\biggl\{
\left(X_{022}^2  + 2  \Cgl X_{022}X_{022}' + \Cgl^2 (
  X_{022}''X_{022} +2
  X_{022}'^{\,2}  ) +\Cgltwo^2X_{022}'^{\,2} \right)  d_{22}^l
\nonumber\\&\hspace{5cm} + \Cgl X_{132}(X_{132} + 2\Cgl X_{132}') d_{33}^l
                + \Cgl X_{121}(X_{121} + 2\Cgl X_{121}') d_{11}^l
\nonumber\\&\hspace{5cm}+   2\Cgltwo\left[ X_{121} X_{132} + \Cgl (X_{121} X_{132}' +
      X_{132} X_{121}')  \right] d_{31}^l + \frac{1}{2}\Cgl^2 X_{220}^2 d_{00}^l
\nonumber\\&\hspace{5cm}+\Cgltwo^2 X_{220}X_{242} d_{40}^l  -2
\Cgl\Cgltwo( X_{220}X_{022}'d_{20}^l +  X_{242}
 X_{022}'d_{42}^l) + \frac{1}{2} \Cgl^2 X_{242}^2 d_{44}^l
\biggr\}
\\
\xil_- &\approx \sum_l \frac{2l+1}{4\pi} (C^E_l -C^B_l)\biggl\{
\left(X_{022}^2 + 2\Cgl X_{022} X_{022}' +
  \Cgl^2(X_{022}''X_{022}+2{X_{022}'}^{2}) +\Cgltwo^2 {X_{022}'}^{2}\right)  d_{2-2}^l
\nonumber\\&\hspace{5cm} +\Cgltwo X_{132}( X_{132} + 2
\Cgl X_{132}') d_{3-3}^l
+\Cgltwo X_{121}(X_{121} + 2 \Cgl X_{121}')d_{1-1}^l
\nonumber\\&\hspace{5cm} +
2 \Cgl\left( X_{121} X_{132} + \Cgl [X_{121} X_{132}' +
      X_{132} X_{121}']  \right) d_{3-1}^l + \frac{1}{2}\Cgltwo^2 X_{220}^2d_{00}^l
\nonumber\\&\hspace{5cm}+ \Cgl^2 X_{220}X_{242} d_{40}^l  -2 \Cgl\Cgltwo (
X_{220}X_{022}'d_{20}^l + X_{242}
 X_{022}'d_{4-2}^l )+ \frac{1}{2}\Cgltwo^2 X_{242}^2 d_{4-4}^l
\biggr\}
\\
\xil_X &\approx \sum_l \frac{2l+1}{4\pi}
C^X_l
\biggl\{ \biggl[X_{022}X_{000} + \Cgl(X_{022}X_{000}' + X_{000}X_{022}') +\Cgltwo^2(X_{000}'X_{022}'
 +\frac{1}{2} X_{220}^2)
\nonumber\\&\hspace{6cm}+\frac{1}{2}\Cgl^2(X_{220}^2 + X_{000}X_{022}'' +
X_{022}X_{000}'' + 4 X_{022}'X_{000}') \biggr]  d_{20}^l
\nonumber\\&\hspace{3cm}+ \frac{2}{\sqrt{l(l+1)}}\left[
  \Cgl(X_{121}X_{000}'' + X_{121}'X_{000}') + X_{121}
  X_{000}'\right]\left( \Cgl d_{1-1}^l +\Cgltwo d_{11}^l\right)
\nonumber\\&\hspace{3cm}
+ \frac{2}{\sqrt{l(l+1)}}\left[
  \Cgl(X_{132}X_{000}'' + X_{132}'X_{000}') + X_{132}
  X_{000}'\right]\left(\Cgltwo d_{3-1}^l + \Cgl d_{31}^l\right)
\nonumber\\&\hspace{3cm} -\Cgl\Cgltwo (X_{000}'X_{220}d_{00}^l + X_{000}'X_{242} d_{40}^l) +
\frac{1}{2} X_{242}X_{220}(\Cgltwo^2d_{4-2}^l+ \Cgl^2d_{42}^l)
\biggl\} .
\end{align}
As discussed in the main text, the $\Cgl$ terms may be neglected at the
$O(10^{-4})$ level for realistic lensing deflection amplitudes.

\end{document}